**Propagation Maps, Maradona Exceptional Points, and Pelé Singularities in Anisotropic, Tellegen, Chiral, Moving-Medium, Omega and Other Isotropy-Broken Materials**

Maxim Durach

Center for Advanced Materials Science, Department of Biochemistry, Chemistry and Physics, Georgia Southern University, Statesboro, GA 30460, USA

Correspondence: mdurach@georgiasouthern.edu

**Abstract** Anisotropic, Tellegen, chiral, moving-medium-type, omega, gyrotropic, hyperbolic, and multi-hyperbolic materials form an important class of isotropy-broken photonic media in which wave propagation can no longer be characterized by the Fresnel wave surface alone. Here we show that Fresnel wave surfaces can be converted into propagation maps that organize positive- and negative-phase-velocity propagation together with attenuation and amplification. In Hermitian media, the boundary between forward and backward propagation forms the Michelangelo silhouette separatrix. This separatrix is also a continuous locus of Maradona exceptional points, where the index-of-refraction operator becomes defective even though the material medium remains Hermitian. In non-Hermitian media, the attenuation-amplification boundary forms the Caravaggio chiaroscuro separatrix. The associated Pelé singularities occur where the handedness remains continuous while the gain-loss character changes sign. Their physical importance is revealed by the momentum-resolved density of states: at these points, the Lorentzian linewidth of the non-Hermitian momentum-resolved density of states (DOS) collapses, producing sharp DOS peaks whose sign reverses across the separatrix. Thus, Pelé singularities are threshold-like gain-loss singularities of the Fresnel wave-surface propagation map, generated by non-Hermitian linewidth collapse. The result is a compact geometric language for describing how handedness, degeneracy, loss, gain, and momentum-resolved DOS are organized in isotropy-broken electromagnetic materials.

## 1. Introduction

The ongoing Fourth Industrial Revolution (4IR) is shaped not only by the rapid emergence of new technologies, but by an increasing unification of the physical, chemical, biological, and digital worlds [1]. In this transformation, electromagnetism plays a special role because it provides one of the most universal languages through which such different domains may be understood, measured, and technologically organized [2-3]. Electricity already serves as a common language of energy in modern civilization, converting diverse fuel sources into transportation, computation, communication, lighting, heating, etc [4]. Metamaterials extend this unifying role further: unlike ordinary materials, whose properties are largely inherited from chemistry alone, metamaterials make structure-property relations explicitly tunable, allowing responses native to one domain of science to be translated into forms accessible to another [5-16]. For this reason, studying different classes of metamaterials is valuable not merely for identifying unusual optical effects, but for establishing the design principles by which various functionalities from different domains of science and technology can be integrated within shared platforms. In this way, metamaterials provide a physical framework through which the converging processes of the Fourth Industrial Revolution can be more consciously understood and directed, so that the technological transformations now unfolding may be guided toward outcomes that more fully serve humanity, protect the natural world, and contribute to the broad aims of the United Nations 2030 Agenda for Sustainable Development, including its 17 Sustainable Development Goals [17].

Modern photonics now encompasses a remarkably broad class of electromagnetic media, including isotropic and anisotropic dielectric, plasmonic, magnetic, gyroelectric and gyromagnetic systems, hyperbolic and multi-hyperbolic materials, and a wide range of bianisotropic responses such as chiral, Tellegen, omega, moving-medium, and axion-type couplings [18-31]. In parallel, contemporary research increasingly extends into non-Hermitian platforms, where loss and gain fundamentally reshape wave propagation [32-35]. In all these cases, the allowed plane-wave states are organized by the Fresnel wave surface, whose geometry encodes the directional structure of propagation in momentum space [35-37]. Over the past decade, Fresnel surfaces have emerged as a central organizing object in photonics, supporting classifications based on topology, asymptotic high-$k$ structure, and singular degeneracies. In particular, the high-$k$ taxonomy introduced by Durach organizes materials into non-, mono-, bi-, tri-, and tetra-hyperbolic phases according to the number of asymptotic double-cone branches [26], while the framework of Favaro and Hehl emphasizes diabolical points at which polarization sheets intersect [38]. At the level of singularities, the standard distinction is between Hermitian diabolical points, where eigenvalues coincide but eigenvectors remain independent, and non-Hermitian exceptional points, where both eigenvalues and eigenvectors coalesce into a defective Jordan structure [39-42].

Despite the power of these classifications, they do not incorporate two physical attributes that are fundamental to electromagnetic wave propagation in complex media. The first is the handedness of the waves and the phase flow relative to energy transport, quantified by the sign of $\boldsymbol{k} \cdot \boldsymbol{S}$, where $\boldsymbol{k}$ is the wavevector and $\boldsymbol{S}$ is the time-averaged Poynting vector. This criterion distinguishes positive-phase-velocity (PPV), negative-phase-velocity (NPV), and orthogonal-phase-velocity (OPV) propagation and becomes especially important in isotropy-broken, bianisotropic, gyrotropic, and multi-hyperbolic systems [43]. The second is the gain-loss character of the waves in non-Hermitian media, encoded in the imaginary part of the refractive index, or equivalently in the sign of $\mathrm{Im}(\boldsymbol{k}) \cdot \boldsymbol{S}$, which distinguishes attenuation from amplification [35, 44]. Existing isofrequency wave-surface classifications consider topology, sheet connectivity, asymptotic cones, and degeneracies, but the organization of the Fresnel wave surface into propagation domains of handedness and gain-loss has not been previously studied.

In this work, we show that Fresnel wave surfaces in Hermitian and non-Hermitian media can be naturally converted into propagation maps. Two scalar functions, $\mathrm{Re}(\boldsymbol{k}) \cdot \boldsymbol{S}$ and $\mathrm{Im}(\boldsymbol{k}) \cdot \boldsymbol{S}$, divide the surface into four distinct sectors corresponding to PPV with loss, PPV with gain, NPV with loss, and NPV with gain. Using the spherical coordinates in momentum space the real part of the refractive index $n$ is positive, and the component of the Poynting vector along the propagation direction, $s$, is positive for PPV ($s > 0$) and negative for NPV ($s < 0$). Writing the complex refractive index as $n \rightarrow n + i\kappa$, where $\kappa = \mathrm{Im}(n)$, the sign of $\kappa$ determines attenuation or amplification through its relation to $s$: $\kappa s > 0$ corresponds to loss, while $\kappa s < 0$ corresponds to gain. The four sectors therefore correspond to PPV with loss ($s > 0, \kappa s > 0$), PPV with gain ($s > 0, \kappa s < 0$), NPV with loss ($s < 0, \kappa s > 0$), and NPV with gain ($s < 0, \kappa s < 0$). This can be seen in Fig. 1(a), where a Fresnel wave surface is depicted with these sectors identified. In this sense, the Fresnel surface acquires a map-like structure: it is partitioned into four neighboring domains of distinct physical behavior, reminiscent of the four-color theorem in planar map coloring, where four colors suffice to distinguish adjacent regions of a map [45]. In this way, the Fresnel wave surface is no longer viewed merely as an isofrequency surface, but as a directional atlas of wave behavior, which is especially important in isotropy-broken media.

The boundaries between these domains are separatrix curves of two different kinds. In Hermitian media, the OPV condition $\boldsymbol{k} \cdot \boldsymbol{S} = 0$ separates forward and backward propagation and defines what we term Michelangelo silhouette separatrices. The purple curves in Fig. 1(a) corresponds to this separatrix. These curves are geometrically natural because, as we demonstrate, they correspond to the silhouette of the Fresnel surface when viewed along $\boldsymbol{k}$, that is, to points where the Fresnel wave-surface normal is orthogonal to the viewing direction from the origin in momentum space. At the same time, they represent the locus where right-handed and left-handed propagation domains meet and touch without crossing. This geometric touching motivates the reference to Michelangelo's *Creation of Adam*, where two opposing hands approach one another at a distinguished boundary between two realities [46]. Because every point on this separatrix marks a touching between the conventional right-handed propagation regime and the left-handed regime, the Michelangelo silhouette separatrix is also a continuous locus of Maradona exceptional points, where the index-of-refraction operator becomes defective even though the material medium remains Hermitian. At these points, the system violates the conventional right-handed propagation rule and touches the left-handed regime, motivating the analogy with the celebrated "Hand of God" goal, where a left-handed touch became a decisive event [47].

In non-Hermitian media, the condition $\mathrm{Im}(\boldsymbol{k}) \cdot \boldsymbol{S} = 0$ separates attenuation from amplification and gives rise to what we term Caravaggio chiaroscuro separatrices. These curves divide dark and bright field domains in a manner analogous to the contrast between shadow and illumination in Caravaggio's paintings [48]. The entire Caravaggio chiaroscuro separatrix is likewise a continuous locus of Pelé singularities, where the phase propagation remains continuous while the field passes from the dark side of the separatrix to the bright side, or vice versa. This behavior is analogous to Pelé's famous no-touch feint, in which the ball continues along essentially the same trajectory while the player crosses sides relative to it without contact [49]. This terminology follows the established precedent of using memorable geometric names for wave and spectral singularities, most notably the term diabolical points, introduced for conical degeneracies whose local double-cone geometry resembles a diabolo. Together, these structures provide a unified framework that incorporates both wave handedness and non-Hermitian gain-loss behavior into the geometry of Fresnel wave surfaces.

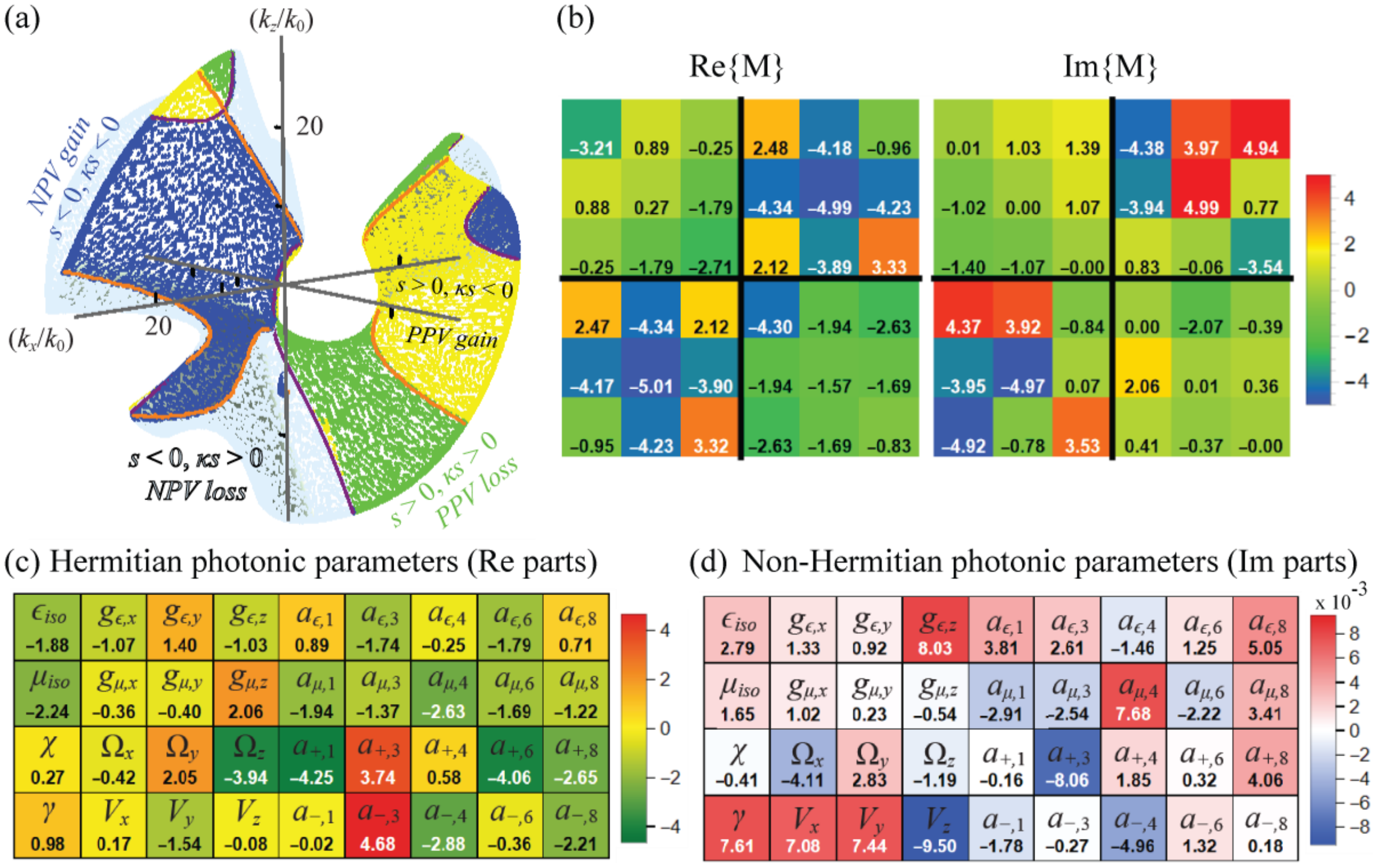

*Figure 1. Fresnel wave surface propagation map and constitutive-parameter decomposition. (a) A Fresnel wave surface divided into four propagation sectors according to the signs of $s$ and $\kappa s$, distinguishing PPV/NPV propagation and loss/gain behavior. The purple curve shows the OPV condition $\boldsymbol{k} \cdot \boldsymbol{S} = 0$, forming the Michelangelo silhouette separatrix between PPV and NPV domains as a continuous locus of Maradona exceptional points. (b) Real and imaginary parts of the constitutive matrix $\widehat{M}$ generating the surface in panel (a). (c,d) Decomposition of $\widehat{M}$ into photonic material parameters: isotropic electric and magnetic response, gyrotropy, anisotropy, Tellegen and chiral response, moving-medium and omega-type magnetoelectric coupling, and symmetric-traceless magnetoelectric anisotropy. The figure shows how the propagation-map geometry of the Fresnel wave surface is traced back to material-response channels.*

To connect these geometric propagation maps to actual material design, we next express the underlying electromagnetic medium through a constitutive matrix and a photonic parameter decomposition. Within the linear regime, diverse classes of materials are described holistically by a constitutive matrix $\widehat{M}$, whose structure encodes the electromagnetic response by relating the fields in electromagnetic waves $(\boldsymbol{D}, \boldsymbol{B})^T = \widehat{M}\ (\boldsymbol{E}, \boldsymbol{H})^T$. Specifically, it is a block matrix

$$\widehat{M} = \begin{pmatrix} \hat{\epsilon} & \hat{X} \\ \hat{Y} & \hat{\mu} \end{pmatrix}, \tag{1}$$

where $\hat{\epsilon}$, $\hat{\mu}$ are the permittivity and permeability, and $\hat{X}$, $\hat{Y}$ are magnetoelectric couplings [23]. In Fig. 1(b) we show the matrix $\widehat{M}$ corresponding to the Fresnel wave surface from Fig. 1(a). Several parametrizations of $\widehat{M}$ have been developed to connect its elements to underlying physical mechanisms. Notable examples include the decomposition of Hehl and Obukhov into principal, skewon, and axion parts [19], and bianisotropy parametrization by Michael Berry [50]. Another important line of work, by Mackay and Lakhtakia, classified constitutive dyadics according to space-time symmetry and magnetic-point-group constraints [23].

In this work, we adopt a novel photonic parametrization, which provides a systematic and physically transparent mapping between the matrix elements of the constitutive tensors and distinct coupling mechanisms. We represent the material tensors $\hat{\epsilon}$, $\hat{\mu}$, $\hat{X}$, and $\hat{Y}$ as

$$\hat{\epsilon} = \epsilon_{iso}\hat{I}_3 + i\big(\boldsymbol{g}_\epsilon \cdot \widehat{\boldsymbol{S}}\big) + \boldsymbol{a}_\epsilon \cdot \widehat{\boldsymbol{A}}, \tag{2}$$

$$\hat{\mu} = \mu_{iso}\hat{I}_3 + i\big(\boldsymbol{g}_\mu \cdot \widehat{\boldsymbol{S}}\big) + \boldsymbol{a}_\mu \cdot \widehat{\boldsymbol{A}} \tag{3}$$

$$\hat{X} = (\chi - i\gamma)\hat{I}_3 + (\boldsymbol{V} + i\boldsymbol{\Omega}) \cdot \widehat{\boldsymbol{S}} + (\boldsymbol{a}_+ - i\boldsymbol{a}_-) \cdot \widehat{\boldsymbol{A}} \tag{4}$$

$$\hat{Y} = (\chi + i\gamma)\hat{I}_3 + (-\boldsymbol{V} + i\boldsymbol{\Omega}) \cdot \widehat{\boldsymbol{S}} + (\boldsymbol{a}_+ + i\boldsymbol{a}_-) \cdot \widehat{\boldsymbol{A}} \tag{5}$$

Here, $\widehat{\boldsymbol{S}} = i(-\lambda_7, \lambda_5, -\lambda_2)$, $\widehat{\boldsymbol{A}} = (\lambda_1, \lambda_3, \lambda_4, \lambda_6, \lambda_8)$, and $\lambda_i$ are Gell-Mann matrices, so that $\widehat{\boldsymbol{S}}$ spans the antisymmetric spin-1 subspace, while $\widehat{\boldsymbol{A}}$ spans the symmetric traceless subspace. The coefficients in this decomposition are obtained by direct projection onto $\lambda_i$ basis matrices and therefore have a clear physical meaning.

The scalars $\epsilon_{iso}$ and $\mu_{iso}$ describe the isotropic electric and magnetic response. The vectors $\boldsymbol{g}_\epsilon$ and $\boldsymbol{g}_\mu$ describe the gyroelectric and gyromagnetic response, respectively. In the magnetoelectric part, $\chi$ is the isotropic Tellegen coupling, $\gamma$ is the isotropic chiral coupling, $\boldsymbol{V}$ is the antisymmetric moving-medium-type nonreciprocal coupling, and $\boldsymbol{\Omega}$ is the antisymmetric omega-type reciprocal magnetoelectric coupling. Finally, $\boldsymbol{a}_\epsilon$ and $\boldsymbol{a}_\mu$ describe the symmetric traceless anisotropic electric and magnetic response, while $\boldsymbol{a}_+$ and $\boldsymbol{a}_-$ describe the symmetric traceless anisotropic Tellegen-type and chiral-type magnetoelectric response, respectively. In Fig. 1(c)–(d), we show the photonic parameters corresponding to the Fresnel wave surface and extracted from matrix $\widehat{M}$ in Fig. 1(b) extracted according to formulas presented in Appendix 1.

One advantage of this parametrization is that it is organized by rotational symmetry. The isotropic terms are SO(3) scalars ($l = 0$), the antisymmetric terms transform as vectors ($l = 1$), and the symmetric traceless terms transform as quadrupolar tensors ($l = 2$). Consequently, the five anisotropy parameters form a real $l = 2$ tensor basis, and their behavior under rotations is governed by the corresponding Wigner rotation matrices. This parametrization allows the propagation-map sectors, Maradona exceptional points, and Pelé singularities introduced in this paper to be traced back to specific material mechanisms: isotropic response, gyrotropy, anisotropy, Tellegen/chiral response, moving-medium coupling, and omega coupling.

## 2. Index of Refraction Operator and Directional Photonics Parameters

To demonstrate the properties of Maradona exceptional points and Pelé singularities, we begin with Maxwell's equations in a linear medium,

$$\nabla \times \boldsymbol{H} - \frac{1}{c}\frac{\partial \boldsymbol{D}}{\partial t} = \frac{4\pi}{c}\boldsymbol{j}_e, -\nabla \times \boldsymbol{E} - \frac{1}{c}\frac{\partial \boldsymbol{B}}{\partial t} = \frac{4\pi}{c}\boldsymbol{j}_m.$$

These equations can be written in operator form as

$$\left(\hat{Q} - \frac{1}{c}\frac{\partial}{\partial t}\widehat{M}\right)\begin{pmatrix}\boldsymbol{E}\\ \boldsymbol{H}\end{pmatrix} = -\frac{4\pi}{c}\begin{pmatrix}\boldsymbol{j}_e\\ \boldsymbol{j}_m\end{pmatrix},$$

with $\hat{Q} = \begin{pmatrix}\hat{0} & \nabla \times \hat{I}\\ -\nabla \times \hat{I} & \hat{0}\end{pmatrix}$. Utilizing the decomposition in Eqs. (2-5) Maxwell's equations for a plane wave with wave vector $\boldsymbol{k}$ and frequency $\omega = k_0 c$ can be written as

$$\hat{\mathcal{L}}(\boldsymbol{k})\begin{pmatrix}\boldsymbol{E}\\ \boldsymbol{H}\end{pmatrix} = \left[\widehat{M}_{iso} \otimes \hat{I}_3 + \sum_{j=1}^{3}\left(\widehat{M}_j^{(S)} + i\frac{k_j}{k_0}\hat{\sigma}_y\right) \otimes \hat{S}_j + \sum_{n=1}^{5}\widehat{M}_n^{(A)} \otimes \hat{A}_n\right]\begin{pmatrix}\boldsymbol{E}\\ \boldsymbol{H}\end{pmatrix} = -\frac{4\pi}{ik_0 c}\begin{pmatrix}\boldsymbol{j}_e\\ \boldsymbol{j}_m\end{pmatrix}.$$

where $\hat{\sigma}_y$ is a Pauli matrix and the nine matrices $\widehat{M}_i$ are

$$\widehat{M}_{iso} = \begin{pmatrix}\epsilon_{iso} & \chi - i\gamma\\ \chi + i\gamma & \mu_{iso}\end{pmatrix}, \qquad M_j^{(S)} = \begin{pmatrix}i\,g_{\epsilon j} & V_j + i\Omega_j\\ -V_j + i\Omega_j & i\,g_{\mu j}\end{pmatrix}, j = 1,2,3,$$

$$M_n^{(A)} = \begin{pmatrix}a_{\epsilon n} & a_{+n} - ia_{-n}\\ a_{+n} + ia_{-n} & a_{\mu n}\end{pmatrix}, n = 1, \dots, 5.$$

To express the field solutions in terms of the sources we use the dyadic Green's function $\hat{G}$

$$\begin{pmatrix}\boldsymbol{E}\\ \boldsymbol{H}\end{pmatrix} = 4\pi i k_0\, \hat{G}\begin{pmatrix}\boldsymbol{j}_e\\ \boldsymbol{j}_m\end{pmatrix}, \quad \hat{G} = i\omega\hat{\mathcal{L}}^{-1} \tag{6}$$

In momentum space, the Green's function is

$$\hat{G}(i\boldsymbol{k}, -ik_0) = i\omega\frac{\operatorname{adj}\hat{\mathcal{L}}(i\boldsymbol{k}, -ik_0)}{\det\hat{\mathcal{L}}(i\boldsymbol{k}, -ik_0)} \tag{7}$$

Once the material response is specified by $\widehat{M}$, the Maxwell operator $\hat{\mathcal{L}}$ determines the dispersion equation for plane waves in the medium. The on-shell wavevectors are those for which the Green's function has poles, or equivalently for which the determinant of the Maxwell operator vanishes [35]. In Hermitian media, these modes are not broadened by loss or gain, so their allowed wavevectors form a sharply defined, or bare, Fresnel wave surface given by

$$F(\boldsymbol{k}) = \det\hat{\mathcal{L}}(i\boldsymbol{k}, -ik_0) = 0. \tag{8}$$

This surface is a two-dimensional manifold in momentum space with zero thickness, corresponding to the delta-functional momentum-resolved 3d density of states $\rho \propto \delta\big(F(\boldsymbol{k})\big)$. We refer to it as a bare Fresnel wave surface to emphasize the absence of non-Hermitian broadening.

To analyze the topology and polarization structure of Fresnel wave surfaces, we employ the index-of-refraction-operator method. Consider a plane wave with fields $\boldsymbol{E}, \boldsymbol{H}, \boldsymbol{D}, \boldsymbol{B} \propto e^{\,i(\boldsymbol{k}\cdot\boldsymbol{r}-k_0 ct)}$, and choose the coordinate system so that the propagation direction is aligned with the $z$-axis, $\boldsymbol{k} = (0,0,k), n = k/k_0$. Maxwell's equations then give

$$\begin{pmatrix}\boldsymbol{D}_\perp\\ \boldsymbol{B}_\perp\end{pmatrix} - n\hat{P}\begin{pmatrix}\boldsymbol{E}_\perp\\ \boldsymbol{H}_\perp\end{pmatrix} = J_4, \qquad \hat{P} = \begin{pmatrix} 0 & 0 & 0 & -1\\ 0 & 0 & 1 & 0\\ 0 & 1 & 0 & 0\\ -1 & 0 & 0 & 0\end{pmatrix}$$

Let the rotated constitutive matrix be block-decomposed into 4x4 block matrix $\widetilde{M}_{\parallel}$, 2x2 matrix $\widetilde{M}_z$, and rectangular matrices $\widetilde{M}_{\parallel z}$ and $\widetilde{M}_{z\parallel}$.

$$\widetilde{M} = \begin{pmatrix}\widetilde{M}_{\parallel} & \widetilde{M}_{\parallel z}\\ \widetilde{M}_{z\parallel} & \widetilde{M}_z\end{pmatrix}.$$

We introduce the 2d constitutive matrix $\widehat{K}$ relating the transverse components of the fields

$$\begin{pmatrix}\boldsymbol{D}_\perp\\ \boldsymbol{B}_\perp\end{pmatrix} = \widehat{K}\begin{pmatrix}\boldsymbol{E}_\perp\\ \boldsymbol{H}_\perp\end{pmatrix}, \quad \text{such that} \quad \{\widehat{K} - n\hat{P}\}\begin{pmatrix}\boldsymbol{E}_\perp\\ \boldsymbol{H}_\perp\end{pmatrix} = J_4 \tag{9}$$

From Maxwell's equations we obtain the relationship between the 3d and 2d constitutive matrices $\widehat{K} = \{\widehat{M}_{\parallel} - \widehat{M}_{\parallel z}\widehat{M}_z^{-1}\widehat{M}_{z\parallel}\}$, and the source vector

$$\hat{J}_4 = \frac{4\pi i}{k_0 c}\left\{\begin{pmatrix}\boldsymbol{j}_{e\perp}\\ \boldsymbol{j}_{m\perp}\end{pmatrix} - \widehat{M}_{\parallel z}\widehat{M}_z^{-1}\begin{pmatrix}j_{ez}\\ j_{mz}\end{pmatrix}\right\}$$

We write

$$\widehat{K} = \begin{pmatrix}\hat{\epsilon}^{(2)} & \hat{X}^{(2)}\\ \hat{Y}^{(2)} & \hat{\mu}^{(2)}\end{pmatrix},$$

and expand each $2\times 2$ block in the natural 2d isotropic - antisymmetric - symmetric-traceless basis of directional photonics parameters using Pauli matrices $\hat{I}_2 = \sigma_0,\ \hat{S}_\perp = -i\sigma_y, \widehat{\boldsymbol{A}}_\perp = (\sigma_z, \sigma_x)$

Thus

$$\hat{\epsilon}^{(2)} = \epsilon_{\text{iso}}^{(2)}\hat{I}_2 + ig_e^{(2)}\hat{S}_\perp + \boldsymbol{a}_e^{(2)}\cdot\widehat{\boldsymbol{A}}_\perp, \qquad \hat{\mu}^{(2)} = \mu_{\text{iso}}^{(2)}\hat{I}_2 + ig_\mu^{(2)}\hat{S}_\perp + \boldsymbol{a}_\mu^{(2)}\cdot\widehat{\boldsymbol{A}}_\perp,$$

$$\hat{X}^{(2)} = \big(\chi^{(2)} - i\gamma^{(2)}\big)\hat{I}_2 + \big(V^{(2)} + i\Omega^{(2)}\big)\hat{S}_\perp + \left(\boldsymbol{a}_+^{(2)} - i\boldsymbol{a}_-^{(2)}\right)\cdot\widehat{\boldsymbol{A}}_\perp,$$

$$\hat{Y}^{(2)} = (\chi^{(2)} + i\gamma^{(2)})\hat{I}_2 + (-V^{(2)} + i\Omega^{(2)})\hat{S}_\perp + (\boldsymbol{a}_+^{(2)} + i\boldsymbol{a}_-^{(2)})\cdot\widehat{\boldsymbol{A}}_\perp.$$

Here $\boldsymbol{a}^{(2)} = \left(a_1^{(2)}, a_2^{(2)}\right)$ are coefficients of a symmetric traceless rank-2 tensor in the transverse plane. The directional photonics parameters – the coefficients in this decomposition - are not global material parameters, but effective parameters associated with the chosen direction $\widehat{\boldsymbol{k}}$. Under a rotation of the transverse axes by angle $\phi$, the scalar and antisymmetric coefficients are invariant, while $\boldsymbol{a}^{(2)}$transforms as

$$\begin{pmatrix} a_1' \\ a_2' \end{pmatrix} = \begin{pmatrix} \cos 2\phi & \sin 2\phi \\ -\sin 2\phi & \cos 2\phi \end{pmatrix} \begin{pmatrix} a_1 \\ a_2 \end{pmatrix}.$$

We rewrite Eq. (9) in terms of the index of refraction operator $\widehat{N} = \widehat{P}^{-1}\left\{\widehat{M}_{\parallel} - \widehat{M}_{\parallel z}\widehat{M}_z^{-1}\widehat{M}_{z\parallel}\right\}$ as

$$\{\widehat{N} - n\hat{I}\}\begin{pmatrix} \boldsymbol{E}_{\perp} \\ \boldsymbol{H}_{\perp} \end{pmatrix} = J_4$$

Following the Refs. [35, 51] we introduce the solution of this problem using the ॐ$_4$-potential

$$\hat{J}_4 = F ॐ_4, \quad \begin{pmatrix} \boldsymbol{E}_{\perp} \\ \boldsymbol{H}_{\perp} \end{pmatrix} = \widehat{U}_4 ॐ_4$$

where $F = \det\{\widehat{N} - n\hat{I}\}$ and $\widehat{U}_4 = \operatorname{adj}\{\widehat{N} - n\hat{I}\}$.

In the absence of sources $\hat{J}_4 = 0$, the generalized eigenproblem $\left\{\widehat{K} - n\widehat{P}\right\}\begin{pmatrix} \boldsymbol{E}_{\perp} \\ \boldsymbol{H}_{\perp} \end{pmatrix} = 0$, converts into ordinary eigenproblem for the index-of-refraction operator $\left\{\widehat{N} - n\hat{I}\right\}\begin{pmatrix} \boldsymbol{E}_{\perp} \\ \boldsymbol{H}_{\perp} \end{pmatrix} = 0$. The operator $\widehat{N}$ can be expressed using the 2d photonic parameters as

$$\widehat{N} = \widehat{N}_{\text{iso}} \otimes \hat{I}_2 + \widehat{N}^{(S)} \otimes \hat{S}_{\perp} + \widehat{N}_x^{(A)} \otimes \sigma_x + \widehat{N}_z^{(A)} \otimes \sigma_z,$$

where

$$\widehat{N}_{\text{iso}} = \begin{pmatrix} V^{(2)} - i\Omega^{(2)} & -ig_{\mu}^{(2)} \\ ig_e^{(2)} & V^{(2)} + i\Omega^{(2)} \end{pmatrix}, \qquad \widehat{N}^{(S)} = \begin{pmatrix} \chi^{(2)} + i\gamma^{(2)} & \mu_{\text{iso}}^{(2)} \\ -\epsilon_{\text{iso}}^{(2)} & -\chi^{(2)} + i\gamma^{(2)} \end{pmatrix},$$

$$\widehat{N}_x^{(A)} = \begin{pmatrix} a_{+,1}^{(2)} + ia_{-,1}^{(2)} & a_{\mu,1}^{(2)} \\ -a_{e,1}^{(2)} & -a_{+,1}^{(2)} + ia_{-,1}^{(2)} \end{pmatrix}, \qquad \widehat{N}_z^{(A)} = \begin{pmatrix} -\left(a_{+,2}^{(2)} + ia_{-,2}^{(2)}\right) & -a_{\mu,2}^{(2)} \\ a_{e,2}^{(2)} & a_{+,2}^{(2)} - ia_{-,2}^{(2)} \end{pmatrix}.$$

Finally, Eq. (8) for the bare Fresnel wave surfaces can be rewritten as [26]

$$F(n) = \det(\widehat{N} - n\hat{I}) = 0, \text{ that is, } n^4 - \operatorname{tr}(\widehat{N})n^3 + \xi n^2 - \zeta n + \det(\widehat{N}) = 0, \text{ with}$$

$$\xi = \frac{1}{2}\left[(\operatorname{tr}\widehat{N})^2 - \operatorname{tr}(\widehat{N}^2)\right], \qquad \zeta = \frac{1}{6}\left[2\operatorname{tr}(\widehat{N}^3) - 3\operatorname{tr}(\widehat{N}^2)\operatorname{tr}(\widehat{N}) + (\operatorname{tr}\widehat{N})^3\right].$$

The four roots of this quartic equation give the refractive-index branches for the chosen direction, and their collection over all directions defines the bare Fresnel wave surface.

### 3. Maradona Exceptional Points and Michelangelo Silhouette Separatrices.

In this manuscript, we show that the coexistence of forward- or positive-phase-velocity and backward- or negative-phase-velocity waves is not a rare pathology, but rather a defining feature of isotropy-broken and multi-hyperbolic media. The boundaries between forward- and backward-wave propagation are the Michelangelo silhouette separatrices, determined by the condition

$$\boldsymbol{k} \cdot \boldsymbol{S} \propto \boldsymbol{k} \cdot \frac{\partial F}{\partial \boldsymbol{k}} = 0,$$

which defines curves on the Fresnel wave surface $F(\boldsymbol{k}) = 0$ separating regions with opposite signs of $\boldsymbol{k} \cdot \boldsymbol{S}$. Mathematically, these curves correspond to the silhouette of the Fresnel wave surface. In differential geometry, silhouette curves are defined as the set of points where the viewing direction is tangent to the surface, or equivalently where the surface normal is orthogonal to the observation direction. In the present case, the surface normal is given by $\nabla_{\boldsymbol{k}} F$, and the condition $\boldsymbol{k} \cdot \nabla_{\boldsymbol{k}} F = 0$ identifies precisely those points where the wavevector lies in the tangent plane to the Fresnel surface. Thus, these separatrices form the silhouette of the surface when viewed along $\boldsymbol{k}$.

Strikingly, these $\boldsymbol{k} \cdot \boldsymbol{S} = 0$ curves are not merely geometric features but correspond to genuine singularities of the index-of-refraction operator. They arise in completely Hermitian, lossless media, yet exhibit the defining algebraic signature of exceptional points: the coalescence of both index-of-refraction eigenvalues and polarization eigenvectors into a defective Jordan structure. This does not contradict the usual association of exceptional points with non-Hermitian systems, because the degeneracy here is not the degeneracy of a Hamiltonian. Rather, it is the degeneracy of the index-of-refraction operator $\widehat{N}$, which generally need not be Hermitian even when the underlying electromagnetic medium is Hermitian.

These Maradona exceptional points differ fundamentally from diabolical points. While both involve eigenvalue degeneracy, diabolical points preserve linearly independent eigenvectors and, therefore, remain diagonalizable. Ordinary diabolical degeneracies require the stronger condition $\frac{\partial F}{\partial \boldsymbol{k}} = 0$, corresponding to vanishing group velocity. In contrast, the Maradona exceptional points identified here occur already under the weaker silhouette condition $\boldsymbol{k} \cdot \frac{\partial F}{\partial \boldsymbol{k}} = 0$, and are therefore directly tied to the transition between forward and backward propagation.

The condition $\boldsymbol{k} \cdot \frac{\partial F}{\partial \boldsymbol{k}} = \boldsymbol{k} \cdot \boldsymbol{S} = 0$ on the Fresnel wave surface corresponds, in the wave-rotated coordinate system, to the double-root condition $F(n) = \frac{dF}{dn} = 0$, and therefore to degenerate refractive-index branches. Indeed, for each direction of propagation, the resonance condition $F(n) = 0$ has four roots $n = n_1, \ldots, n_4$ corresponding to the four indices of refraction of the respective electromagnetic waves. The necessary and sufficient condition for a coalescence of two roots at $n_1 = n_2 = n_0$ is

$$F(n) = 0 \quad \text{and} \quad \frac{dF}{dn} = 4n^3 - 3\mathrm{tr}(\widehat{N})\, n^2 + 2\xi \mathrm{n} - \zeta = 0$$

For a simple root $F(n) = (n - n_0)F'(n)$ with $F'(n_0) \neq 0$. However, for a double root $F(n) = (n - n_0)^2 F''(n)/2$ with $F''(n_0) \neq 0$, but $F'(n_0) = 0$.

Thus, the Michelangelo silhouette condition is not only a geometric condition on the Fresnel wave surface, but also the algebraic condition for coalescence of refractive-index eigenvalues. When this root coalescence is accompanied by eigenvector coalescence of the generally non-Hermitian operator $\widehat{N}$, the result is a Maradona exceptional point in a Hermitian electromagnetic medium.

As an initial illustration, let us consider the eigenvectors $(\boldsymbol{E}_\perp, \boldsymbol{H}_\perp)$ corresponding to the coalescent roots $n_1 = n_2 = n_0$. Assume first that both $\boldsymbol{E}_\perp$ and $\boldsymbol{H}_\perp$ are linearly polarized in the transverse $xy$-plane. On the Michelangelo silhouette separatrix, $\boldsymbol{k} \cdot \boldsymbol{S} = 0$, so the longitudinal component of the Poynting vector vanishes. For linearly polarized transverse fields, this means that the directions of $\boldsymbol{E}_\perp$ and $\boldsymbol{H}_\perp$ coincide. By rotating the transverse coordinate system around $\widehat{\boldsymbol{k}}$, this common transverse direction can be aligned with the $x$-axis. In that frame, the full $\boldsymbol{E}$ and $\boldsymbol{H}$ fields lie in the $xz$-plane, and the Abraham momentum axis is aligned with the $y$-axis, so that $\boldsymbol{S} \parallel \widehat{\boldsymbol{y}}$. The eigenvector can then be written as $(\boldsymbol{E}_\perp, \boldsymbol{H}_\perp) = (Z_\perp, 0,1,0)$, where $Z_\perp = \frac{E_x}{H_x}$ is the transverse impedance. This impedance must be the same for both waves associated with the degenerate roots $n_1 = n_2 = n_0$. Indeed, it satisfies the linear equation $Z_\perp \widehat{N}_1 + \widehat{N}_3 = n_0\, (Z_\perp, 0,1,0)^T$, where $\widehat{N}_1$ and $\widehat{N}_3$ are the first and third columns of $\widehat{N}$. This equation admits at most one value of $Z_\perp$. Therefore, although the eigenvalue $n_0$ is doubly degenerate, the corresponding eigenspace is at most one-dimensional. The two roots $n_1 = n_2 = n_0$ thus share the same eigenvector, implying eigenvector coalescence and defectiveness of $\widehat{N}$ in an exceptional-point-like manner, but without requiring material loss or gain.

Consider a Hermitian reciprocal material with $\hat{\epsilon} = \mathrm{diag}\{-2, -1, 1\}$, $\hat{\mu} = \mathrm{diag}\{1, 2, -1\}$. This material has twelve diabolical singularities, as can be seen from the Fresnel wave surface of this material shown in Fig. 2(a). One of them, marked by a red dot, corresponds to the direction $\widehat{\boldsymbol{k}} = \left(\sqrt{3/5}, 0, \sqrt{2/5}\right)$. The index-of-refraction operator for this direction is diagonalizable

$$\widehat{N} = \begin{pmatrix} 0 & B \\ C & 0 \end{pmatrix}, \qquad B = \begin{pmatrix} 0 & 2 \\ 5 & 0 \end{pmatrix}, \qquad C = \begin{pmatrix} 0 & 1 \\ \frac{5}{2} & 0 \end{pmatrix}.$$

Its eigenvalues are $n_i = \pm\sqrt{5}$ and it possesses four independent eigenvectors, $(\boldsymbol{E}_\perp, \boldsymbol{H}_\perp) = \left(\pm 2, 0, 0, \sqrt{5}\right)$, $(0, \pm\sqrt{5}, 1, 0)$. Thus, this is an ordinary diabolical degeneracy: the eigenvalues are degenerate, but the eigenspace remains complete.

By contrast, this is not true for the Maradona exceptional points on the Michelangelo silhouette separatrix. In Fig. 2(a), the green and light-blue regions show the two handedness domains on the bare Fresnel wave surface, with green denoting $s > 0$ PPV propagation and light blue denoting $s < 0$ NPV propagation. For the anisotropic material considered here, these PPV and NPV regions are associated with the electric-like and magnetic-like hyperbolic branches respectively identified in Ref. [52]. The purple curves show the Michelangelo silhouette separatrices.

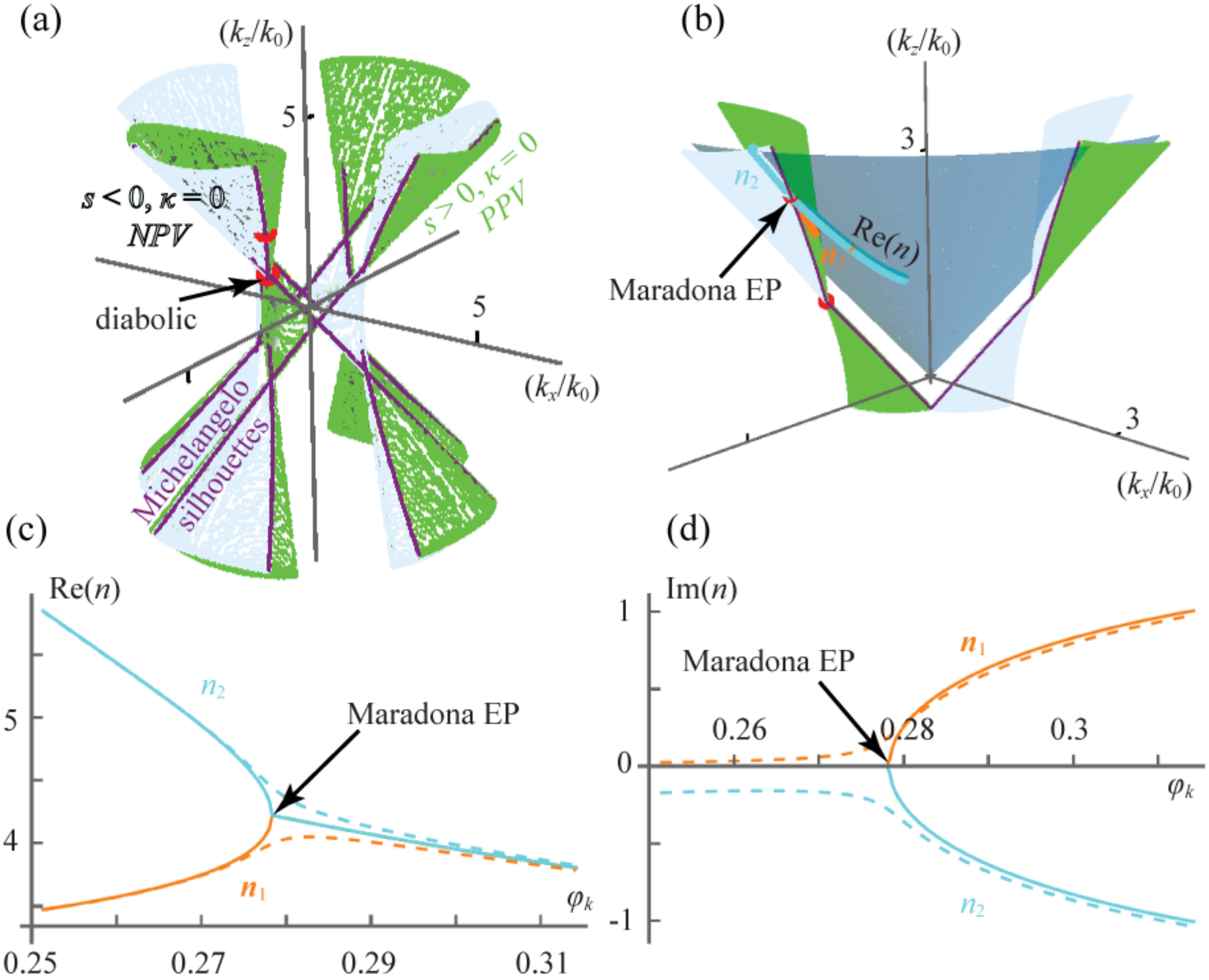

*Figure 2. Michelangelo silhouette separatrices and Maradona exceptional points. (a) Bare Fresnel wave surface for the Hermitian reciprocal material* $\hat{\varepsilon} = \mathrm{diag}(-2,-1,1)$, $\hat{\mu} = \mathrm{diag}(1,2,-1)$. *The colored regions indicate PPV (green) and NPV (light blue) domains, while the purple curves show the Michelangelo silhouette separatrices defined by* $\boldsymbol{k} \cdot \boldsymbol{S} = 0$. *Diabolical point and Maradona EP discussed in the text are indicated by red dots. (b) Close up of the Maradona exceptional point. The red point and black arrow mark the Maradona EP, and the blue cone represents nearby propagation directions used to scan the refractive-index branches. The traces on the cone show* $\mathrm{Re}(n_1)$ *(cyan) and* $\mathrm{Re}(n_2)$ *(orange). (c) Real parts of the two refraction index branches as functions of the angular direction* $\varphi_k$ *along the cone in (b). The indices of refraction coalesce at the Maradona EP. The dashed lines show the unfolding of the Maradona EP when a small loss perturbation is added* $\widehat{M} \to \widehat{M} + \delta\widehat{M}$, $\delta\widehat{M} = 10^{-3} i\, \hat{I}_6$. *(d) Imaginary parts of the same branches, showing the exceptional-point-type unfolding of the refractive-index eigenvalues. The medium is Hermitian, but the degeneracy occurs in the generally non-Hermitian index-of-refraction operator* $\widehat{N}$.

In Fig. 2 we show the Michelangelo silhouette separatrices using the purple curves. Consider, for example, the direction $\hat{\boldsymbol{k}} = \left(\frac{7}{\sqrt{107}}, \frac{2}{\sqrt{107}}, \frac{3\sqrt{6}}{\sqrt{107}}\right)$ and its intersection with Fresnel wave surface also shown using a red dot. The corresponding index-of-refraction operator is

$$\hat{N} = \begin{pmatrix} 0 & B \\ C & 0 \end{pmatrix}, \quad B = \begin{pmatrix} -\sqrt{\frac{214}{3}} & -\frac{113}{3} \\ 0 & \sqrt{\frac{214}{3}} \end{pmatrix}, \qquad C = \begin{pmatrix} -\frac{\sqrt{\frac{107}{6}}}{2} & -\frac{79}{24} \\ 0 & \frac{\sqrt{\frac{107}{6}}}{2} \end{pmatrix}.$$

This operator is defective. Its Jordan normal form is

$$\hat{N}_J = \begin{pmatrix} J_- & 0 \\ 0 & J_+ \end{pmatrix}, J_\pm = \begin{pmatrix} \pm n_0 & 1 \\ 0 & \pm n_0 \end{pmatrix}, n_0 = \frac{\sqrt{107}}{\sqrt{6}}.$$

The $4 \times 4$ operator $\hat{N}$ therefore has only two ordinary eigenvectors $\mathbf{v} = (\boldsymbol{E}_\perp, \boldsymbol{H}_\perp)$, with

$$\boldsymbol{H}_\perp = \left(-\frac{5\sqrt{2}}{321}, 1\right), \boldsymbol{E}_\perp = \pm 2\, \boldsymbol{H}_\perp, \qquad \text{and } Z_\perp = \pm 2.$$

The corresponding generalized eigenvectors are

$$\mathbf{u} = (\boldsymbol{E}_\perp, \boldsymbol{H}_\perp) = \left(\frac{12}{107}, \frac{110}{7}\sqrt{\frac{2}{321}}, \pm\frac{424}{2247}, 0\right).$$

The red point in Fig. 2(b) shows this Maradona exceptional point on Fresnel wave surface. The blue cone defines a one-parameter family of nearby directions. Moving along this cone gives the angular coordinate $\varphi_k$ used in Figs. 2(c) and 2(d). The real parts of the two refractive-index branches are traced directly on the cone in Fig. 2(b) and then plotted as functions of $\varphi_k$ in Fig. 2(c). At the Maradona EP these two real branches coalesce, consistent with the double-root condition $F(n) = dF/dn = 0$. When a small non-Hermitian perturbation is added, the degeneracy unfolds: the real parts separate and the imaginary parts shown in Fig. 2(d) become nonzero. This illustrates the sensitivity of the Maradona EP to loss or gain perturbations. While the perturbation is small $\delta\hat{M} = 10^{-3} i\, \hat{I}_6$, the splitting is orders of magnitude larger $|\delta n| \propto 0.1$.

In the Jordan-degenerate case, the index-of-refraction operator $\hat{N}$ admits an eigenvector $\mathbf{v}$ and a generalized eigenvector $\mathbf{u}$ satisfying

$$(\hat{N} - \lambda I)\mathbf{v} = 0, \qquad (\hat{N} - \lambda I)\mathbf{u} = \mathbf{v}.$$

The corresponding field solution takes the form of a Maradona mode,

$$\psi(z) = e^{ik_0\hat{N}z}\psi(0) = [(c_1 + ic_2 k_0 z)\mathbf{v} + c_2\mathbf{u}]e^{ik_0 n_0 z}.$$

This expression exhibits the characteristic linear-in-$z$ prefactor associated with Jordan-block dynamics, formally analogous to the behavior at exceptional points. However, since $\boldsymbol{k} \cdot \boldsymbol{S} \neq 0$ for the generalized eigenvector $\mathbf{u}$, the condition $\boldsymbol{k} \cdot \boldsymbol{S} = 0$ is violated if $c_2 \neq 0$. In an unbounded Hermitian medium, where $n_0$ is real, the condition $c_2 = 0$ is therefore formally enforced; otherwise, the linear prefactor would produce unbounded amplitude growth, in contrast to Voigt-type solutions in non-Hermitian media [53].

Thus, although Maradona singularities have the algebraic structure of exceptional points, they arise within Hermitian electromagnetic media and are intrinsically linked to the Michelangelo silhouette separatrices, where forward- and backward-propagating domains meet. This gives the physical reason for Maradona exceptional points. At OPV, $\boldsymbol{k} \cdot \boldsymbol{S} = 0$, so the transverse electric and magnetic fields no longer form the oriented-area structure that normally produces longitudinal Poynting flux. In the simplest case, they become parallel in the transverse plane. At the same time, OPV marks a turning point of the refractive-index branches for a fixed propagation direction. Thus, the loss of longitudinal energy flow is accompanied by a loss of polarization and impedance distinction between the coalescing modes, allowing the index-of-refraction operator $\widehat{N}$ to become defective at the Maradona exceptional point.

## 4. Pelé Singularities and Caravaggio Chiaroscuro Separatrices

We now extend the propagation-map construction to non-Hermitian media. We introduce loss or gain by perturbing the Hermitian constitutive matrix $\widehat{M}$ as

$$\widehat{M}' = \widehat{M} + \delta\widehat{M},$$

where $\widehat{M} = \widehat{M}^{\dagger}$ describes the unperturbed lossless material and $\delta\widehat{M}$ represents the non-Hermitian part of the response. In the photonic-parameter representation introduced above, this corresponds to allowing the material parameters to acquire complex corrections. The Maxwell operator is modified accordingly, $\hat{\mathcal{L}}' = \hat{Q} + i\omega\widehat{M}'$, and the refractive index becomes complex,

$$n \to n + i\kappa.$$

The coefficient $\kappa$ describes extinction or amplification and the most direct way to locate the boundary between these two behaviors is to impose $\kappa = 0$. At this boundary the refractive index is real, even though the medium is non-Hermitian. Therefore, the non-Hermitian dispersion equation $\det\hat{\mathcal{L}}' = 0$ must admit a real-$k$ solution. Since $\det\hat{\mathcal{L}}'$ is generally complex, this gives the two real conditions

$$\mathrm{Re}\{\det\hat{\mathcal{L}}'\} = 0,\ \mathrm{Im}\{\det\hat{\mathcal{L}}'\} = 0. \tag{10}$$

This real/imaginary decomposition of a complex quartic dispersion condition was previously used in the inverse problem of quartic photonics, where complex bianisotropic media were represented through the intersection of analogous real and imaginary quartic surfaces [54]. In that work, the construction provided a way to relate prescribed complex photonic states to effective material parameters. Here, we use the same operational principle in a different context: the simultaneous solution of Eqs. (10) identifies real-$k$ filaments on the non-Hermitian Fresnel wave surface. We interpret these filaments as Caravaggio chiaroscuro separatrices, because they mark the loci where the complex refractive index becomes real and the gain-loss character of the propagation map changes. The orange curve in Fig. 1(a) corresponds to this type of separatrix.

The physical distinction between attenuation and amplification is not determined by the sign of $\kappa$ alone, but by the sign of its product with the energy-flow component along the propagation direction, $\kappa S_z = \kappa s$. For a given propagation branch, $\kappa s > 0$ corresponds to attenuation, while $\kappa s < 0$ corresponds to amplification. Thus, the Caravaggio separatrix divides the non-Hermitian propagation map into dark and bright domains, corresponding respectively to decaying and amplified fields. The zeros of $\kappa$ therefore define the boundaries at which the gain-loss character reverses. We refer to these $\kappa = 0$ curves as Caravaggio chiaroscuro separatrices, because they divide dark and bright field domains in a manner analogous to the contrast between shadow and illumination in Caravaggio's paintings.

These properties are illustrated in Fig. 3, where the momentum-resolved DOS distribution $\rho$ is shown for the $k_z = 0$ cross-section of the Fresnel wave surface from Fig. 1(a). The direction-dependent sign of the density of states, sign $\rho(\hat{\mathbf{k}})$, is an important characteristic of the material. It indicates which waves transfer energy from external sources into the medium, corresponding to $\rho(\hat{\mathbf{k}}) > 0$ (absorption), and which correspond to the opposite direction of energy transfer, $\rho(\hat{\mathbf{k}}) < 0$ (gain) [35,55]. In Fig. 3(a) the intersections of the curves $\mathrm{Re}\{\det \hat{\mathcal{L}}'\} = 0$, $\mathrm{Im}\{\det \hat{\mathcal{L}}'\} = 0$ are marked by black arrows and numbered 1–4. These intersections identify the real-$k$ Caravaggio separatrices and coincide with the sign changes in the DOS.

The singular points associated with these $\kappa = 0$ curves are termed Pelé singularities. At $\kappa = 0$, the phase-propagation direction remains continuous, while the gain-loss character, determined by the sign of $\kappa S_z$, reverses across the Caravaggio separatrix. This is analogous to Pelé's famous no-touch feint: the ball continues along essentially the same trajectory, while the player passes across that trajectory without contact, switching sides relative to it. Likewise, at a Pelé singularity, the phase propagation remains continuous, but the field passes from the dark side of the separatrix to the bright side, or vice versa. In this sense, Pelé singularities play for the non-Hermitian gain-loss map a role analogous to that played by Maradona exceptional points for the Hermitian handedness switching.

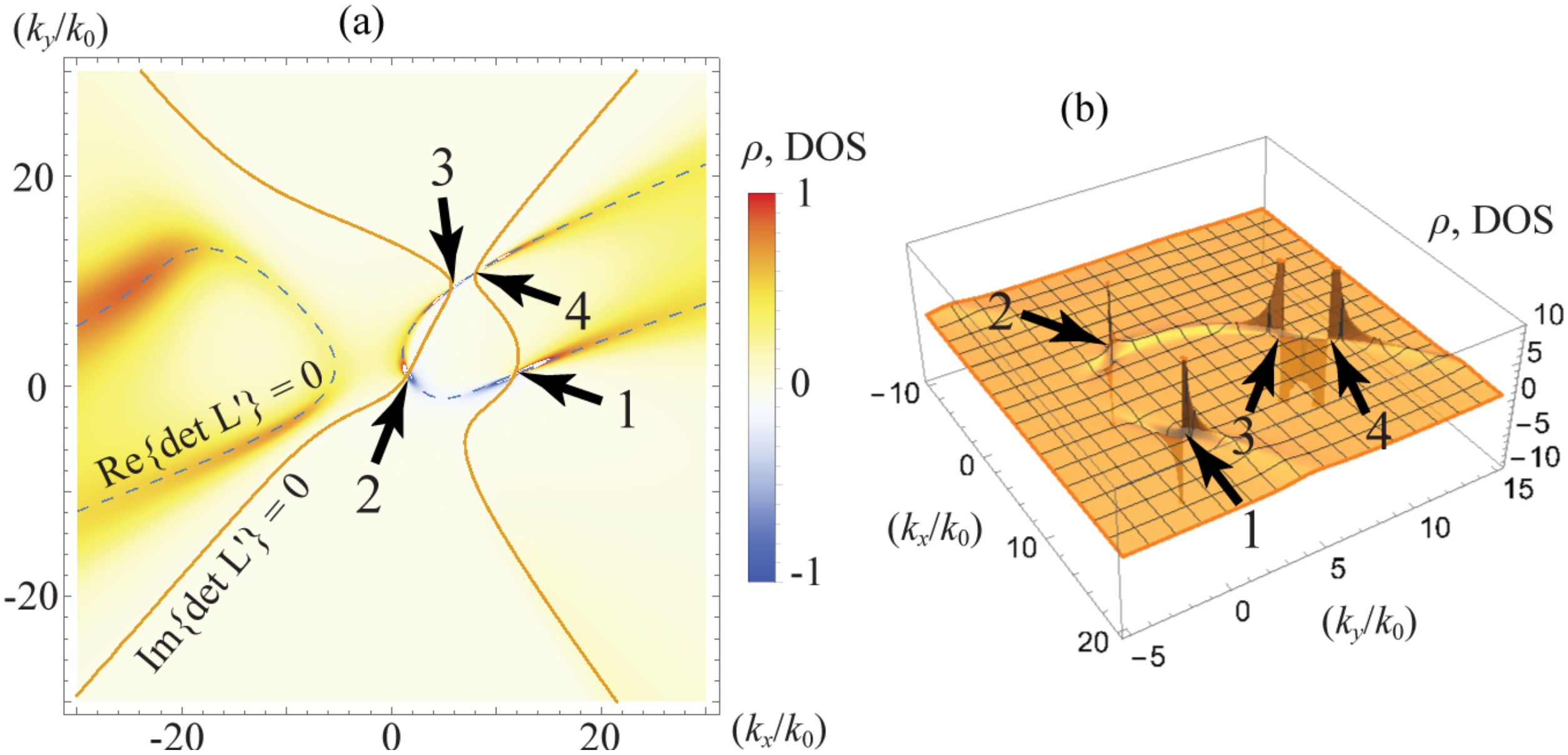

*Figure 3. Caravaggio chiaroscuro separatrices and momentum-resolved density of states in a non-Hermitian Fresnel wave surface map. (a) Momentum-resolved density of states ρ in the $k_z = 0$ cross-section of the non-Hermitian Fresnel wave surface. The orange and dashed curves show the real and imaginary dispersion conditions,* $\mathrm{Re}\,\{\det\hat{\mathcal{L}}'\} = 0$ *and* $\mathrm{Im}\,\{\det\hat{\mathcal{L}}'\} = 0$. *Their intersections, marked by numbered arrows, define real-$k$ filaments where $\kappa = 0$, corresponding to Caravaggio chiaroscuro separatrices. Across these curves the gain-loss character changes sign. (b) Three-dimensional view of the same DOS landscape. The peaks and sign changes illustrate how Pelé singularities appear as gain-loss reversal points in the broadened momentum-resolved density of states.*

The same condition can also be expressed perturbatively in terms of the fields. Expanding the determinant of the non-Hermitian Maxwell operator to first order gives

$$\det\hat{\mathcal{L}}' = \det\,(\hat{\mathcal{L}} + \delta\hat{\mathcal{L}}) = \det\,\hat{\mathcal{L}} + \mathrm{tr}\big[\mathrm{adj}\,(\hat{\mathcal{L}})\delta\hat{\mathcal{L}}\big].$$

On the unperturbed Fresnel surface, $\det\hat{\mathcal{L}} = 0$, so the non-Hermitian correction is governed by the projection of the perturbation onto the unperturbed mode. If $\mathbf{v} = (\boldsymbol{E}, \boldsymbol{H})$ is the eigenvector of the corresponding Hermitian mode, this correction may be written as $\mathrm{tr}\big[\mathrm{adj}\,(\hat{\mathcal{L}})\delta\hat{\mathcal{L}}\big] \propto \mathbf{v}^T\delta\widehat{M}\mathbf{v}$.

Equivalently, introducing the complex energy $U \propto \boldsymbol{E}\cdot\boldsymbol{D}^* + \boldsymbol{H}\cdot\boldsymbol{B}^*$ through $\mathrm{Im}(k)\,S_z = \mathrm{Im}(U), \mathrm{Re}(k)\,S_z = \mathrm{Re}(U)$, one obtains the quality factor of the electromagnetic wave as

$$Q^{-1} = \frac{\mathrm{Im}(k)}{\mathrm{Re}(k)} = \frac{\mathrm{Im}(U)}{\mathrm{Re}(U)}.$$

Using $k = k_0 n$, this gives

$$\kappa = n\,\frac{\mathrm{Im}(U)}{\mathrm{Re}(U)} = \frac{\mathrm{Im}(U)}{S_z}.$$

For a weak non-Hermitian perturbation $\delta\widehat{M}$

$$\delta n = i\kappa = n\,\frac{\mathbf{v}^T \delta\widehat{M}\,\mathbf{v}}{\mathbf{v}^T \widehat{M}\,\mathbf{v}}$$

Thus, the condition $\mathbf{v}^T \delta\widehat{M}\mathbf{v} = 0$ is the perturbative field-based form of the more operational Caravaggio chiaroscuro separatrix condition Eq. (10). In this sense, the determinant equations locate the separatrix directly, while the modal expression explains how a particular electromagnetic mode experiences the non-Hermitian perturbation. Together with the Michelangelo silhouette separatrix $\boldsymbol{k}\cdot\boldsymbol{S} = 0$, the Caravaggio chiaroscuro separatrix completes the propagation map: one boundary separates PPV from NPV propagation, while the other separates attenuation from amplification.

## 5. Momentum-Resolved Density of States and Maradona-Pelé Singularities

The addition of loss or gain gives the refractive-index branches imaginary parts

$$n_i \to n_i + i\kappa_i.$$

As a result, the delta-functional DOS of the bare Fresnel wave surface is broadened (see Fig. 3 for illustration). In the Hermitian limit, the momentum-resolved density of states is concentrated on shell. In the non-Hermitian case, this delta function is replaced by a Lorentzian [35],

$$\rho(\widehat{\boldsymbol{k}}) = \sigma\,\delta(k - k_0 n) \to \rho = \frac{\sigma}{\pi}\,\frac{k_0\kappa}{(k - k_0 n)^2 + (k_0\kappa)^2}.$$

Here $\sigma = \sigma(\widehat{\boldsymbol{k}}), n = n(\widehat{\boldsymbol{k}}), \kappa = \kappa(\widehat{\boldsymbol{k}})$ are direction-dependent quantities: $\sigma$ is the surface-density-of-states prefactor, $n$ is the real part of the refractive index, and $\kappa$ is the extinction/amplification coefficient. From the Lorentzian form above the sign of the broadened DOS can be expressed as

$$\operatorname{sign}\rho = \operatorname{sign}\sigma\,\operatorname{sign}\kappa. \tag{11}$$

Following Appendix 2, the surface-density prefactor $\sigma$ can be written as

$$\sigma = \frac{6\omega^3}{c^3} n_0^3 \left(\frac{E_x^2}{\boldsymbol{E}\cdot\boldsymbol{D}^* + \boldsymbol{H}\cdot\boldsymbol{B}^*}\right) = \frac{6\omega^2}{c^2} n_0^3 \left(\frac{E_x^2}{2\boldsymbol{k}\cdot\mathrm{Re}\{\boldsymbol{E}\times\boldsymbol{H}^*\}}\right)$$

The sign of SDOS $\operatorname{sign}\sigma = \operatorname{sign}(\boldsymbol{k}\cdot\boldsymbol{S})$, which means that for right-handed waves $\sigma > 0$, while for left-handed waves $\sigma < 0$. Relation Eq. (11) help combine the two sign structures of the propagation map: the factor $\sigma$ carries the handedness information through $\boldsymbol{k}\cdot\boldsymbol{S}$, while $\kappa$ carries the information about the gain-loss due to the non-Hermitian perturbation

It is noteworthy that, similarly to $\sigma$, the sign of the extinction coefficient $\kappa$ is governed by the sign of $\boldsymbol{k}\cdot\boldsymbol{S} \propto \mathbf{v}^T \widehat{M}\,\mathbf{v}$. As a result, $\kappa$ changes sign upon crossing a Maradona exceptional point, so that modes on opposite sides of a Michelangelo silhouette separatrix possess opposite extinction. Consequently, according to Eq. (11) the sign of the density of states $\rho$ and loss-gain state remains unchanged across Michelangelo silhouettes and only changes at Pelé singularities

$$\operatorname{sign}\rho = \operatorname{sign}\kappa s. \tag{12}$$

The defining feature of Maradona exceptional points is $\boldsymbol{k}\cdot\boldsymbol{S} = 0$, which leads to divergence of the surface density of states (SDOS) $\sigma \propto (\boldsymbol{k}\cdot\boldsymbol{S})^{-1} \to \infty$. Nevertheless, the significance of this is reduced by the fact that for Hermitian media this divergence adds to already divergent delta-functions momentum-resolved DOS distribution. In non-Hermitian media, this divergence is generally unfolded by perturbations of the Maradona EP, as illustrated in Fig. 2.

Pelé singularities produce a more dramatic DOS signature. Since Pelé singularities occur at $\kappa = 0$, they correspond to points where the Lorentzian linewidth $k_0\kappa$ of the momentum-resolved spectral response collapses. At the on-shell momentum $k = k_0 n$, the Lorentzian amplitude scales as

$$\rho(k = k_0 n) \sim \frac{\sigma}{\pi k_0 \kappa},$$

so the DOS develops a sharp peak whose sign reverses across the Pelé singularity, as illustrated in Fig. 3(b), where four Pelé singularities are marked by black arrows. Thus, a Pelé singularity is not only a boundary between attenuation and amplification; it is a threshold-like momentum-space singularity of the non-Hermitian Fresnel wave surface map. Its origin stems from the vanishing of the non-Hermitian linewidth of the dressed Fresnel wave surface. In this sense, Pelé singularities identify directions in momentum space where the exchange of energy between the field and the medium reverses sign.

## 6. Conclusion

We have shown that Fresnel wave surfaces can be interpreted as propagation maps that organize both handedness and gain-loss behavior. In Hermitian media, the Michelangelo silhouette separatrices $\boldsymbol{k}\cdot\boldsymbol{S} = 0$ separates PPV and NPV domains and supports Maradona exceptional points, where the index-of-refraction operator becomes defective even though the material medium remains Hermitian. In non-Hermitian media, the Caravaggio chiaroscuro separatrix is defined by real-$k$ solutions of the complex dispersion equation,

$$\operatorname{Re}\{\det\hat{\mathcal{L}}'\} = 0, \qquad \operatorname{Im}\{\det\hat{\mathcal{L}}'\} = 0,$$

and marks the reversal between attenuation and amplification. The associated Pelé singularities occur where the handedness remains continuous while the gain-loss character changes sign. Their physical importance is amplified by the momentum-resolved density of states: at $\kappa = 0$, the Lorentzian linewidth of the non-Hermitian momentum-resolved response collapses, producing a sharp peak in momentum-resolved DOS whose sign reverses across the separatrix, as illustrated in Fig. 3(b). Thus, Pelé singularities are threshold-like gain-loss singularities of the Fresnel wave surface propagation map, generated by non-Hermitian linewidth collapse. Together, Michelangelo/Maradona and Caravaggio/Pelé structures provide compact geometric language for describing how handedness, degeneracy, loss, gain, and momentum-resolved DOS are organized on Fresnel wave surfaces.

**Appendix 1. Photonic Material Parameters Retrieval**

Using Eqs. (2)-(5) we obtain the isotropic parts as

$$\epsilon_{iso} = \frac{\mathrm{tr}(\hat{\epsilon})}{3}, \qquad \mu_{iso} = \frac{\mathrm{tr}(\hat{\mu})}{3}, \qquad \chi = \frac{1}{3}\,\mathrm{tr}\left(\frac{\hat{X}+\hat{Y}}{2}\right), \qquad \gamma = \frac{1}{3}\,\mathrm{tr}\left(\frac{i(\hat{X}-\hat{Y})}{2}\right).$$

Thus, $\epsilon_{iso}$ and $\mu_{iso}$ describe the isotropic electric and magnetic response, while $\chi$ and $\gamma$ are the isotropic Tellegen and chiral magnetoelectric couplings, respectively.

The antisymmetric parts of $\hat{\epsilon}$ and $\hat{\mu}$, projected onto the spin-1 basis $\boldsymbol{S}$, define the gyroelectric and gyromagnetic vectors $\boldsymbol{g}_{\epsilon}$ and $\boldsymbol{g}_{\mu}$:

$$(g_{\epsilon})_i = -\frac{1}{2}\,\mathrm{tr}[(-iS_i)\,\hat{\epsilon}], \qquad (g_{\mu})_i = -\frac{1}{2}\,\mathrm{tr}[(-iS_i)\hat{\mu}], \qquad i = 1,2,3.$$

These three-component vectors encode the antisymmetric electric and magnetic response sectors.

For the magnetoelectric tensor combinations the antisymmetric spin-1 projections define the vectors $\boldsymbol{V}$ and $\boldsymbol{\Omega}$:

$$V_i = -\frac{1}{2}\,\mathrm{tr}\left[(-iS_i)\left(\frac{\hat{X}+\hat{Y}}{2}\right)\right], \qquad \Omega_i = -\frac{1}{2}\,\mathrm{tr}\left[(-iS_i)\left(\frac{i(\hat{X}-\hat{Y})}{2}\right)\right], \qquad i = 1,2,3.$$

In this parametrization, $\boldsymbol{V}$ describes the nonreciprocal moving-medium-type coupling, whereas $\boldsymbol{\Omega}$ corresponds to the reciprocal omega-type magnetoelectric coupling.

The remaining anisotropic contributions are contained in the symmetric traceless coefficients $\boldsymbol{a}_{\epsilon}, \boldsymbol{a}_{\mu}, \boldsymbol{a}_{+}, \boldsymbol{a}_{-}$, obtained by projection onto the five-element basis $\mathbf{A}$:

$$(a_e)_j = \frac{1}{2}\,\mathrm{tr}(A_j\hat{\epsilon}), \qquad (a_{\mu})_j = \frac{1}{2}\,\mathrm{tr}(A_j\hat{\mu}), \qquad (a_+)_j = \frac{1}{2}\,\mathrm{tr}\left(A_j\left(\frac{\hat{X}+\hat{Y}}{2}\right)\right),$$

$$(a_-)_j = \frac{1}{2}\,\mathrm{tr}\left(A_j\left(\frac{i(\hat{X}-\hat{Y})}{2}\right)\right), \qquad j = 1,\ldots,5.$$

These coefficients describe the symmetric traceless anisotropic parts of the electric, magnetic, and magnetoelectric response tensors.

In this way, the full constitutive response is decomposed into three parts: an isotropic part proportional to $\hat{I}_3$, an antisymmetric spin-1 part expanded in $\boldsymbol{S}$, and a symmetric traceless part expanded in $\boldsymbol{A}$.

**Appendix 2. Surface Density of States (SDOS)**

Consider operator $\hat{\Sigma}(\lambda)=\frac{\hat{L}}{ik}=\left(\frac{\hat{Q}}{k}+\lambda\hat{M}\right)$, where $\lambda=1/n$ . Near Fresnel wave surface $\lambda\approx\lambda_0$ we expand

$$\hat{\Sigma}(\lambda)=\hat{\Sigma}(\lambda_0)+(\lambda-\lambda_0)\hat{M}+\cdots$$

$$\hat{\Sigma}^{-1}(\lambda)=\frac{\hat{\Sigma}_{-1}}{\lambda-\lambda_0}+\hat{\Sigma}_0+\cdots$$

$$\hat{1}=\hat{\Sigma}(\lambda)\,\hat{\Sigma}^{-1}(\lambda)=\left\{\hat{\Sigma}(\lambda_0)+(\lambda-\lambda_0)\hat{M}+\cdots\right\}\left\{\frac{\hat{\Sigma}_{-1}}{\lambda-\lambda_0}+\hat{\Sigma}_0+\cdots\right\}$$

From this $\hat{\Sigma}(\lambda_0)\hat{\Sigma}_{-1}=0$, since the left-hand side has no pole at Fresnel wave surface. This means that each column of $\hat{\Sigma}_{-1}$ is a multiple of the nullvector $\mathbf{v}$ of $\hat{\Sigma}(\lambda_0)$: $\hat{\Sigma}_{-1}=\mathbf{v}\otimes\boldsymbol{\alpha}^T$, where $\boldsymbol{\alpha}$ is some unknown vector. This results in

$$\hat{\Sigma}(\lambda_0)\hat{\Sigma}_0+\hat{M}\hat{\Sigma}_{-1}=\hat{1}$$

Multuplying by the left nullvector $\mathbf{w}^T$ and using $\mathbf{w}^T\hat{\Sigma}(\lambda_0)=0$, we get $\mathbf{w}^T\hat{M}(\mathbf{v}\otimes\boldsymbol{\alpha}^T)=\boldsymbol{w}^T$, or

$$\left(\mathbf{w}^T\hat{M}\,\mathbf{v}\right)\boldsymbol{\alpha}^T=\mathbf{w}^T,\quad \boldsymbol{\alpha}^T=\frac{\mathbf{w}^T}{\mathbf{w}^T\hat{M}\,\mathbf{v}}.$$

Finally,

$$\hat{\Sigma}^{-1}(\lambda)=\frac{\mathbf{v}\otimes\mathbf{w}^T}{(\lambda-\lambda_0)\left(\mathbf{w}^T\hat{M}\,\mathbf{v}\right)}$$

For Hermitian materials $w=v=(\boldsymbol{E},\boldsymbol{H})$, therefore,

$$L^{-1}=-\frac{ikn_0^2}{(n-n_0)}\left(\frac{\mathbf{v}\otimes\mathbf{v}^T}{\mathbf{v}^T\hat{M}\,\mathbf{v}}\right)=-ik_0\frac{n_0^3}{(n-n_0)}\left(\frac{\mathbf{v}\otimes\mathbf{v}^T}{\mathbf{v}^T\hat{M}\,\mathbf{v}}\right)$$

From this

$$\rho=\frac{6\omega}{\pi c^2}\operatorname{Im}\{\hat{G}_{xx}\}=\frac{6\omega^2}{\pi c^2}\operatorname{Re}\left\{\left(\hat{L}^{-1}\right)_{xx}\right\}=\frac{6\omega^3}{\pi c^3}n_0^3\left(\frac{E_x^2}{\boldsymbol{E}\cdot\boldsymbol{D}^*+\boldsymbol{H}\cdot\boldsymbol{B}^*}\right)\operatorname{Im}\left\{\frac{1}{n-n_0}\right\}$$

The resulting expression for the prefactor $\sigma$ is

$$\sigma=\frac{6\omega^3}{c^3}n_0^3\left(\frac{E_x^2}{\boldsymbol{E}\cdot\boldsymbol{D}^*+\boldsymbol{H}\cdot\boldsymbol{B}^*}\right)=\frac{6\omega^2}{c^2}n_0^3\left(\frac{E_x^2}{2\boldsymbol{k}\cdot\operatorname{Re}\{\boldsymbol{E}\times\boldsymbol{H}^*\}}\right)$$